\documentclass[10pt,conference]{IEEEtran}
\IEEEoverridecommandlockouts
\makeatletter
\def\ps@headings{%
\def\@oddhead{\mbox{}\scriptsize\rightmark \hfil \thepage}%
\def\@evenhead{\scriptsize\thepage \hfil \leftmark\mbox{}}%
\def\@oddfoot{}%
\def\@evenfoot{}}
\makeatother 

\IEEEoverridecommandlockouts
\usepackage{bbm}
\usepackage{amsfonts}
\usepackage[dvips]{graphicx}
\usepackage{times}
\usepackage{cite}
\usepackage{amsmath}
\usepackage{array}
\usepackage{amssymb}
\usepackage{siunitx}

\usepackage{stfloats}
\usepackage{diagbox}
\usepackage{graphicx}

\usepackage{footnote}
\usepackage{microtype}
\usepackage{booktabs}
\usepackage{array}
\usepackage{algorithm}
\usepackage{subeqnarray}
\usepackage{cases}
\usepackage{threeparttable}
\usepackage{color}
\usepackage{hyperref}
\usepackage{epstopdf}
\usepackage{algpseudocode}
\usepackage{bm}
\usepackage{multirow}
\usepackage[labelformat=simple]{subcaption}
\usepackage{adjustbox}

\makeatletter
\usepackage[dvipsnames]{xcolor}
\PassOptionsToPackage{dvipsnames,svgnames,table}{xcolor}%
\makeatother

\usepackage{listofitems}
\usepackage{pgffor}
\usepackage{xparse}
\usepackage{ifthen}

\usepackage{pgfplots}
  \pgfplotsset{compat=newest}
  \usetikzlibrary{plotmarks}
  \usetikzlibrary{arrows.meta}
  \usepgfplotslibrary{patchplots}
  \usepackage{grffile}
  \usepackage{amsmath}
\usepackage{nicefrac}
\usepgfplotslibrary{groupplots,dateplot}
\pgfplotsset{compat=1.18}
\usepgfplotslibrary{fillbetween}
\usepackage{tikz}
\usetikzlibrary{positioning}
\usetikzlibrary{patterns}
\usepgfplotslibrary{polar}
\usetikzlibrary{shapes}
\usetikzlibrary{external}
\usepackage[acronym,shortcuts]{glossaries}
\makeglossaries

\newacronym{2d}{2D}{two-dimensional}
\newacronym{3d}{3D}{three-dimensional}
\newacronym{3gpp}{3GPP}{3rd generation partnership project}
\newacronym{3msk}{3MSK}{3-level minimum-shift-keying}
\newacronym{4g}{4G}{fourth-generation}
\newacronym{5g}{5G}{fifth-generation}
\newacronym{6dof}{6DoF}{six degrees of freedom}
\newacronym{6g}{6G}{sixth-generation}
\newacronym{abp}{ABP}{authentication by personalisation}
\newacronym{ac}{AC}{alternating current}
\newacronym{aci}{ACI}{adjacent channel interference}
\newacronym{aclr}{ACLR}{adjacent channel leakage ratio}
\newacronym{acpr}{ACPR}{adjacent channel power ratio}
\newacronym{adc}{ADC}{analog-to-digital converter}
\newacronym{adr}{ADR}{adaptive data rate}
\newacronym{aec}{AEC}{aluminum electrolytic capacitor}
\newacronym{aes}{AES}{Advanced Encryption Standard}
\newacronym{afe}{AFE}{analog front end}
\newacronym{ag}{AG}{array gain}
\newacronym{agc}{AGC}{automatic gain controller}
\newacronym{agps}{A-GPS}{Assisted Global Positioning System} 
\newacronym{ai}{AI}{artificial intelligence}
\newacronym{alk}{ALK}{Alkaline}
\newacronym{am}{AM}{amplitude modulation}
\newacronym{amam}{AM/AM}{amplitude modulation to amplitude modulation}
\newacronym{amc}{AMC}{automatic modulation classification}
\newacronym{amp}{AMP}{approximate message passing}
\newacronym{ampm}{AM/PM}{amplitude modulation to phase modulation}
\newacronym{aoa}{AOA}{angle-of-arrival}
\newacronym{aod}{AOD}{angle-of-departure}
\newacronym{ap}{AP}{access point}
\newacronym{api}{API}{application program interface}
\newacronym{apt}{APT}{acoustic power transfer}
\newacronym{apu}{APU}{access point unit}
\newacronym{ar}{AR}{augmented reality}
\newacronym{arima}{ARIMA}{Auto-Regressive Integrated Moving Average}
\newacronym{arp}{ARP}{Antenna Reference Point}
\newacronym{asic}{ASIC}{application specific integrated circuit}
\newacronym{ask}{ASK}{amplitude-shift keying}
\newacronym{at}{AT}{ATtention}
\newacronym{auv}{AUV}{autonomous underwater vehicle}
\newacronym{awg}{AWG}{arbitrary waveform generator}
\newacronym{awgn}{AWGN}{additive white Gaussian noise}
\newacronym{baw}{BAW}{bulk acoustic wave}
\newacronym{bb}{BB}{base-band}
\newacronym{bc}{BC}{backscatter communication}
\newacronym{bcjr}{BCJR}{Bahl-Cocke-Jelinek-Raviv}
\newacronym{bd}{BD}{backscatter device}
\newacronym{be}{BE}{Belgium}
\newacronym{ber}{BER}{bit error rate}
\newacronym{bf}{BF}{beamforming}
\newacronym{bga}{BGA}{ball grid array}
\newacronym{bldc}{BLDC}{brushless DC}
\newacronym{bler}{BLER}{block error rate}
\newacronym{bod}{BOD}{Brown-Out Detection}
\newacronym{bom}{BOM}{bill of materials}
\newacronym{bpsk}{BPSK}{binary phase-shift keying}
\newacronym{brp}{BRP}{beam refinement process}
\newacronym{bs}{BS}{base station}
\newacronym{bu}{BU}{booster unit}
\newacronym{bw}{BW}{bandwidth}
\newacronym{ca}{CA}{carrier emitter}
\newacronym{cad}{CAD}{channel activity detection}
\newacronym{cars}{CARS}{calibration reference signal}
\newacronym{cb}{CB}{conjugate beamforming}
\newacronym{cbm}{CBM}{condition based maintenance}
\newacronym{cc}{CC}{constant current}
\newacronym{ccdf}{CCDF}{complementary cumulative distribution function}
\newacronym{ccnn}{CCNN}{circular convolutional neural network}
\newacronym{ccs}{CCS}{correlative channel sounder}
\newacronym{cdf}{CDF}{cumulative distribution function}
\newacronym{cdma}{CDMA}{code division-multiple access}
\newacronym{cdrx}{CDRX}{connected mode DRX}
\newacronym{ce}{CE}{coverage enhancement}
\newacronym{ced}{CED}{cumulative energy density}
\newacronym{ceofdm}{CE-OFDM}{constant-envelope OFDM}
\newacronym{cf}{CF}{cell-free}
\newacronym{cfmmimo}{CF-mMIMO}{cell-free massive MIMO}
\newacronym{cfo}{CFO}{carrier frequency offset}
\newacronym{cir}{CIR}{channel impulse response}
\newacronym{cla}{CLA}{closed-loop approach}
\newacronym{clk}{CLK}{clock}
\newacronym{cmos}{CMOS}{complementary metal oxide semiconductor}
\newacronym{cnn}{CNN}{convolutional neural network}
\newacronym{co}{CO}{combinatorial optimization}
\newacronym{cordic}{CORDIC}{coordinate rotation digital computer}
\newacronym{cost}{COST}{commercial off-the-shelf}
\newacronym{cots}{COTS}{commercial off-the-shelf}
\newacronym{cp}{CP}{cyclic prefix}
\newacronym{cpe}{CPE}{common phase error}
\newacronym{cpfsk}{CPFSK}{continuous phase frequency shift keying}
\newacronym{cpm}{CPM}{continuous phase modulation}
\newacronym{cpt}{CPT}{capacitive power transfer}
\newacronym{cpu}{CPU}{central-processing unit}
\newacronym{cpw}{CPW}{coplanar waveguide}
\newacronym{cqi}{CQI}{channel quality indicator}
\newacronym{cr}{CR}{coding rate}
\newacronym{crc}{CRC}{cyclic redundancy check}
\newacronym{crlb}{CRLB}{Cram\'er-Rao lower bound}
\newacronym{crs}{CRS}{cell reference signal}
\newacronym{cs}{CS}{compressed sensing}
\newacronym{cse}{CSE}{channel state estimation}
\newacronym{csi}{CSI}{channel state information}
\newacronym{csp}{CSP}{contact service point}
\newacronym{css}{CSS}{chirp spread spectrum}
\newacronym{cu}{CU}{central unit}
\newacronym{cv}{CV}{constant voltage}
\newacronym{cw}{CW}{continuous wave}
\newacronym{d2d}{D2D}{device-to-device}
\newacronym{dab}{DAB}{Digital Audio Broadcasting}
\newacronym{dac}{DAC}{digital-to-analog converter}
\newacronym{daq}{DAQ}{data acquisition system}
\newacronym{das}{DAS}{distributed antenna systems}
\newacronym{dbpsk}{DBPSK}{Differential Binary Phase Shift Keying}
\newacronym{dc}{DC}{direct current}
\newacronym{dcc}{DCC}{dynamic cooperation clustering}
\newacronym{ddc}{DDC}{digital down conversion}
\newacronym{de}{DE}{drain efficiency}
\newacronym{dft}{DFT}{discrete Fourier transform}
\newacronym{dftsofdm}{DFT-s-OFDM}{discrete Fourier Transform spread OFDM}
\newacronym{dftsofdmfdss}{DFT-s-OFDM-FDSS}{DFT-s-OFDM with frequency-domain spectral shaping}
\newacronym{dftsofdmfdssse}{DFT-s-OFDM-FDSS-SE}{DFT-s-OFDM with FDSS and spectral extension}
\newacronym{dl}{DL}{downlink}
\newacronym{dlc}{DLC}{Distributed Laser Charging}
\newacronym{dli}{DLI}{direct link interference}
\newacronym{dlt}{DLT}{Distributed Ledger Technology}
\newacronym{dm}{DM}{diffuse multipath}
\newacronym{dma}{DMA}{Direct Memory Access}
\newacronym{dmac}{DMAC}{Direct Memory Access Controller}
\newacronym{dmc}{DMC}{diffuse multipath component}
\newacronym{dmimo}{D-MIMO}{distributed MIMO}
\newacronym{dnn}{DNN}{deep neural network}
\newacronym{doa}{DOA}{direction-of-arrival}
\newacronym{dp}{DP}{differencial privacy}
\newacronym{dpd}{DPD}{digital pre-distortion}
\newacronym{dpdk}{DPDK}{Data Plane Development Kit}
\newacronym{dram}{DRAM}{dynamic random-access memory}
\newacronym{drcs}{$\Delta$RCS}{differential-radar cross section}
\newacronym{drx}{DRX}{Discontinuous Reception Mode}
\newacronym{dsb}{DSB}{double-sideband}
\newacronym{dsl}{DSL}{digital subscriber line}
\newacronym{dsp}{DSP}{digital signal processing}
\newacronym{dss}{DSS}{Dataset Storage Standard}
\newacronym{duc}{DUC}{digital up-converter}
\newacronym{dvb}{DVB}{Digital Video Broadcasting}
\newacronym{e2e}{E2E}{end-to-end}
\newacronym{easa}{EASA}{European Union Aviation Safety Agency}
\newacronym{ebg}{EBG}{electromagnetic bandgap}
\newacronym{ebw}{EBW}{excess bandwidth}
\newacronym{ec}{EC}{European Commission}
\newacronym{ecc}{ECC}{elliptic curve cryptography}
\newacronym{ecdf}{eCDF}{empirical cumulative distribution function}
\newacronym{ecdlp}{ECDLP}{elliptic curve discrete logarithm problem}
\newacronym{ecsp}{ECSP}{edge computing service point}
\newacronym{edlc}{EDLC}{electrostatic double-layer capacitors}
\newacronym{edrx}{eDRX}{Extended Discontinuous Reception Mode}
\newacronym{ee}{EE}{energy efficiency}
\newacronym{egprs}{EGPRS}{Enhanced Data Rates for GSM Evolution}
\newacronym{eh}{EH}{energy harvesting}
\newacronym{eirp}{EIRP}{equivalent isotropically radiated power}
\newacronym{em}{EM}{electromagnetic}
\newacronym{embb}{eMBB}{enhanced Mobile Broadband}
\newacronym{en}{EN}{energy neutral}
\newacronym{end}{END}{energy-neutral device}
\newacronym{enob}{ENOB}{effective number of bits}
\newacronym{eol}{EoL}{end of life}
\newacronym{ep}{EP}{energy profiler}
\newacronym{epd}{EPD}{electronic paper display}
\newacronym{epu}{EPU}{edge processing unit}
\newacronym{er}{ER}{Energy Receiver}
\newacronym{erp}{ERP}{effective radiation power}
\newacronym[plural=ESCs,firstplural=electronic speed controllers (ESCs)]{esc}{ESC}{electronic speed control}
\newacronym{esd}{ESD}{electrostatic discharge}
\newacronym{esl}{ESL}{electronic shelf label}
\newacronym{esr}{ESR}{equivalent series resistance}
\newacronym{et}{ET}{Energy Transmitter}
\newacronym{etsi}{ETSI}{European Telecommunications Standards Institute}
\newacronym{evd}{EVD}{eigenvalue decomposition}
\newacronym{evm}{EVM}{Error Vector Magnitude}
\newacronym{ewlb}{eWLB}{Embedded Wafer Level Ball Grid Array}
\newacronym{fa}{FA}{federation anchor}
\newacronym{fair}{FAIR}{Findability, Accessibility, Interoperability, and Reuse of digital assets}
\newacronym{fcf}{FCF}{frequency correlation function}
\newacronym{fd}{FD}{front-haul distance}
\newacronym{fdd}{FDD}{frequency-division duplexing}
\newacronym{fde}{FDE}{frequency domain equalizer}
\newacronym{fdm}{FDM}{frequency-division multiplexing}
\newacronym{fdma}{FDMA}{frequency division multiple access}
\newacronym{fdss}{FDSS}{frequency-domain spectral shaping}
\newacronym{fem}{FEM}{finite element analysis}
\newacronym{fembb}{feMBB}{further enhanced mobile broadband}
\newacronym{fft}{FFT}{fast Fourier transform}
\newacronym{fh}{FH}{fronthaul}
\newacronym{fhss}{FHSS}{frequency hopping spread spectrum}
\newacronym{fifo}{FIFO}{First In, First Out}
\newacronym{fim}{FIM}{Fisher information matrix}
\newacronym{fits}{FITS}{Flexible Image Transport System}
\newacronym{fl}{FL}{federated learning}
\newacronym{fom}{FoM}{figure of merit}
\newacronym{fov}{FOV}{field of view}
\newacronym{fpga}{FPGA}{field-programmable gate array}
\newacronym{fr1}{FR1}{frequency range 1}
\newacronym{fr2}{FR2}{frequency range 2}
\newacronym{fram}{FRAM}{Ferroelectric Random Access Memory}
\newacronym{fsk}{FSK}{frequency shift keying}
\newacronym{fspl}{FSPL}{free space path loss}
\newacronym{fss}{FSS}{frequency selective surface}
\newacronym{gb}{GB}{grant-based}
\newacronym{gdpr}{GDPR}{general data protection regulation}
\newacronym{gf}{GF}{grant-free}
\newacronym{gmsk}{GMSK}{Gaussian minimum-shift keying}
\newacronym{gnb}{gNB}{Next Generation Node B}
\newacronym{gni}{GNI}{gross national income}
\newacronym{gnn}{GNN}{graph neural network}
\newacronym{gnss}{GNSS}{global navigation satellite system}
\newacronym{gpclk}{GPCLK}{general purpose clock}
\newacronym{gpio}{GPIO}{General-Purpose Input/Output}
\newacronym{gpl}{GPL}{GNU General Public License}
\newacronym{gprs}{GPRS}{General Packet Radio Services}
\newacronym{gps}{GPS}{Global Positioning System}
\newacronym{gpu}{GPU}{graphical processing unit}
\newacronym{grc}{GRC}{GNU Radio Companion}
\newacronym{gscm}{GSCM}{geometry‐based stochastic model}
\newacronym{gsm}{GSM}{Global System for Mobile Communications}  %
\newacronym{gspm}{GSpM}{Generalized spatial modulation}
\newacronym{gwp}{GWP}{Global Warming Potential}
\newacronym{harq}{HARQ}{hybrid automatic repeat request}
\newacronym{hat}{HAT}{hardware attached on top}
\newacronym{hcs}{HCS}{human-centric services}
\newacronym{hdf5}{HDF5}{Hierarchical Data Format version 5}
\newacronym{hfss}{HFSS}{High Frequency Simulator Software}
\newacronym{hmd}{HMD}{head-mounted display}
\newacronym{hpbm}{HPBM}{half power beam width}
\newacronym{HyMPRo}{HyMPRo}{Hybrid Multi-Path Routing algorithm}
\newacronym{i.i.d.}{i.i.d.}{independent and identically distributed}
\newacronym{i2c}{I2C}{Inter-Integrated Circuit}
\newacronym{iaq}{IAQ}{Indoor Air Quality}
\newacronym{ib}{IB}{in-band}
\newacronym{ibbc}{IBBC}{inter-band beam configuration}
\newacronym{ibo}{IBO}{input back-off}
\newacronym{ic}{IC}{integrated circuit}
\newacronym{iccs}{ICCS}{Ilmsens correlative channel sounder}
\newacronym{ici}{ICI}{intercarrier interference}
\newacronym{icnirp}{ICNIRP}{International Commission on Non-Ionizing Radiation Protection}
\newacronym{id}{ID}{information decoding}
\newacronym{idft}{IDFT}{inverse discrete Fourier transform}
\newacronym{idxm}{IDXM}{index modulation}
\newacronym{if}{IF}{intermediate-frequency}
\newacronym{ifft}{IFFT}{inverse fast-Fourier-transform}
\newacronym{iid}{i.i.d.}{independently and identically distributed}
\newacronym{iis}{IIS}{integrated information system}
\newacronym{im}{IM}{intermodulation}
\newacronym{imd}{IMD}{intermodulation distortion}
\newacronym{imu}{IMU}{inertial measurement unit}
\newacronym{inh}{InH}{indoor hotspot office}
\newacronym{io}{IO}{input/output}
\newacronym{ioe}{IoE}{Internet of Everything}
\newacronym{iot}{IoT}{Internet of Things}
\newacronym{ipt}{IPT}{inductive power transfer}
\newacronym{ipy}{IPY}{Interventions per Year}
\newacronym{iq}{IQ}{in-phase and quadrature}
\newacronym{iqi}{IQI}{IQ imbalance}
\newacronym{ir}{IR}{infrared}
\newacronym{isi}{ISI}{intersymbol interference}
\newacronym{ism}{ISM}{industrial, scientific and medical}
\newacronym{isp}{ISP}{internet service provider}
\newacronym{itu}{ITU}{International Telecommunication Union}
\newacronym{jesd}{JESD}{Joint Electron Devices Engineering Council}
\newacronym{jfet}{JFET}{junction field effect transistor}
\newacronym{kpi}{KPI}{key performance indicator}
\newacronym{ktofdm}{KT-DFT-s-OFDM}{known-tail-DFT-s-OFDM}
\newacronym{kvi}{KVI}{key value indicator}
\newacronym{larva}{LARVA}{LARge Virtual Array}
\newacronym{lca}{LCA}{life cycle assessment}
\newacronym{lco}{LCO}{lithium cobalt oxide}
\newacronym{ldo}{LDO}{Low-dropout voltage regulator}
\newacronym{ldpc}{LDPC}{low-density parity-check}
\newacronym{led}{LED}{Light Emitting Diode}
\newacronym{less}{LESS}{Low Energy Scheduler Solution}
\newacronym{lfp}{LFP}{lithium iron phosphate}
\newacronym{lib}{LIB}{Lithium-Ion Battery}
\newacronym{lic}{LIC}{lithium-ion capacitor}
\newacronym{lid}{LID}{Lithium Iron Disulfide}
\newacronym{lidar}{LiDAR}{light detection and ranging}
\newacronym{liion}{Li-ion}{lithium-ion}
\newacronym{lipo}{LiPo}{lithium polymer}
\newacronym{lis}{LIS}{large intelligent surface}
\newacronym{llh}{LLH}{log-likelihood}
\newacronym{lls}{LLS}{link-level simulation}
\newacronym{lmd}{LMD}{Lithium Manganese Dioxide}
\newacronym{lmmse}{LMMSE}{least minimum mean square error}
\newacronym{lmo}{LMO}{lithium ion manganese oxide}
\newacronym{lna}{LNA}{low-noise amplifier}
\newacronym{lo}{LO}{local oscillator}
\newacronym{lora}{LoRa}{long range}
\newacronym{lorawan}{LoRaWAN}{long-range wide-area network}
\newacronym{los}{LoS}{line-of-sight}
\newacronym{lp}{LP}{linear programming}
\newacronym{lpf}{LPF}{low-pass filter}
\newacronym{lpt}{LPT}{laser power transfer}
\newacronym{lpwa}{LPWA}{Low Power Wide Area}
\newacronym{lpwan}{LPWAN}{low-power wide-area network}
\newacronym{lpwans}{LPWANs}{Low-Power Wide-Area Networks}
\newacronym{lqi}{LQI}{link quality indicator}
\newacronym{lrelu}{LReLU}{leaky rectified linear unit}
\newacronym{lrt}{LRT}{likelihood-ratio test}
\newacronym{ls}{LS}{least squares}
\newacronym{lsa}{LSA}{large synthetic array}
\newacronym{lsf}{LSF}{large-scale fading}
\newacronym{lsfc}{LSFC}{large-scale fading component}
\newacronym{lstm}{LSTM}{Long Short-Term Memory}
\newacronym{ltc}{LTC}{lithium thionyl chloride}
\newacronym{lte}{LTE}{Long Term Evolution}
\newacronym{lti}{LTI}{linear time-invariant}
\newacronym{lto}{LTO}{lithium titanate}
\newacronym{lusta}{LUSTA 5G }{Logistique mUltimodale Sécuritaire Téléopérée \& Autonome 5G}
\newacronym{m2m}{M2M}{machine to machine}
\newacronym{mac}{MAC}{Medium Access Control}
\newacronym{mate}{MATE}{millimeter-wave MIMO testbed}
\newacronym{mc}{MC}{Monte Carlo}
\newacronym{mcl}{MCL}{Maximum Coupling Loss}
\newacronym{mcs}{MCS}{modulation and coding scheme}
\newacronym{mcu}{MCU}{microcontroller unit}
\newacronym{mec}{MEC}{multi-access edge computing}
\newacronym{mems}{MEMS}{micro-electromechanical systems}
\newacronym{mf}{MF}{matched filter}
\newacronym{mimo}{MIMO}{multiple-input multiple-output}
\newacronym{miso}{MISO}{multiple-input single-output}
\newacronym{ml}{ML}{machine learning}
\newacronym{mlp}{MLP}{multilayer perceptron}
\newacronym{mmic}{MMIC}{monolithic microwave integrated circuit}
\newacronym{mmimo}{mMIMO}{massive MIMO}
\newacronym{mmse}{MMSE}{minimum mean square error}
\newacronym{mmtc}{mMTC}{massive machine-typed communication}
\newacronym{mmwave}{mmWave}{millimeter wave}
\newacronym{mn}{MN}{matching network}
\newacronym{mosfet}{MOSFET}{metal-oxide semiconductor field effect transistor}
\newacronym{mpc}{MPC}{multipath component}
\newacronym{mppt}{MPPT}{maximum power point tracking}
\newacronym{mr}{MR}{maximum ratio}
\newacronym{mrc}{MRC}{maximum ratio combining}
\newacronym{mrc_em}{MRC}{maximum ratio combining}
\newacronym{mrc_EM}{MRC}{Magnetic Resonance Coupling}
\newacronym{mrt}{MRT}{maximum ratio transmission}
\newacronym{mse}{MSE}{mean square error}
\newacronym{msk}{MSK}{Minimum-Shift Keying}
\newacronym{mtc}{MTC}{Machine-Type Communication}
\newacronym{multi-rat}{Multi-RAT}{multiple radio access technology}
\newacronym{multirat}{Multi-RAT}{Multiple Radio Access Technology}
\newacronym{music}{MUSIC}{MUltiple SIgnal Classification}
\newacronym{navauwall}{NAVAUWALL}{AUtomated NAVigation in WALLonia}
\newacronym{nb}{NB}{narrowband}
\newacronym{nbiot}{NB-IoT}{narrowband IoT}
\newacronym{nca}{NCA}{nickel cobalt aluminum}
\newacronym{netcdf}{NetCDF}{Network Common Data Form}
\newacronym{nf}{NF}{noise figure}
\newacronym{nfc}{NFC}{near-field communication}
\newacronym{nfv}{NFV}{network function virtualization}
\newacronym{ngmn}{NGMN}{Next Generation Mobile Networks }
\newacronym{ni}{NI}{National Instruments}
\newacronym{nicd}{NiCd}{nikkel cadmium}
\newacronym{nimh}{NiMH}{nikkel metal hydride}
\newacronym{nlos}{NLoS}{non-line-of-sight}
\newacronym{nmc}{NMC}{nickel manganese cobalt}
\newacronym{nmos}{nMOS}{n-channel metal-oxide semiconductor}
\newacronym{nn}{NN}{neural network}
\newacronym{nnls}{NNLS}{non-negative least squares}
\newacronym{noma}{NOMA}{non-orthogonal multiple access}
\newacronym{np}{NP}{Neyman-Pearson}
\newacronym{npbch}{NPBCH}{Narrowband Physical Broadcast Channel}
\newacronym{npss}{NPSS}{Narrow Band Primay Synchronization Signal}
\newacronym{nr}{NR}{New Radio}
\newacronym{nrs}{NRS}{Narrow Band Reference Signal}
\newacronym{nsss}{NSSS}{Narrowband Secondary Synchronization Signal}
\newacronym{ntp}{NTP}{network time protocol}
\newacronym{oai}{OAI}{OpenAirInterface} 
\newacronym{obw}{OBW}{occupied bandwidth}
\newacronym{ofdm}{OFDM}{orthogonal frequency-division multiplexing}
\newacronym{ofdma}{OFDMA}{orthogonal frequency-division multiple access}
\newacronym{ofdmim}{OFDM-IM}{OFDM with index modulation}
\newacronym{olos}{OLoS}{obstructed-line-of-sight}
\newacronym{oma}{OMA}{orthogonal multiple access}
\newacronym{oob}{OOB}{out-of-band}
\newacronym{ook}{OOK}{on-off keying}
\newacronym{opbo}{OPBO}{output power backoff}
\newacronym{oran}{O-RAN}{open radio-access network}
\newacronym{os}{OS}{operating system}
\newacronym{ota}{OTA}{over-the-air}
\newacronym{otaa}{OTAA}{over-the-air authentication}
\newacronym{p1}{P1}{Phase 1}
\newacronym{p2}{P2}{Phase 2}
\newacronym{p2p}{P2P}{point-to-point}
\newacronym{pa}{PA}{power amplifier}
\newacronym{pae}{PAE}{power-added efficiency}
\newacronym{pam}{PAM}{pulse amplitude modulation}
\newacronym{pana}{PanA}{Panel A}
\newacronym{panb}{PanB}{Panel B}
\newacronym{papr}{PAPR}{peak-to-average power ratio}
\newacronym{pc}{PC}{pilot count}
\newacronym{pcb}{PCB}{printed circuit board}
\newacronym{pcg}{PCG}{power consumption gain}
\newacronym{pcie}{PCIe}{Peripheral Component Interconnect Express}
\newacronym{pcsi}{PCSI}{perfect channel state information}
\newacronym{pd}{PD}{powered device}
\newacronym{pdcch}{PDCCH}{physical downlink control channel}
\newacronym{pdf}{PDF}{probability density function}
\newacronym{pdp}{PDP}{power delay profile}
\newacronym{pdsch}{PDSCH}{physical downlink shared channel}
\newacronym{pe}{PE}{processing element}
\newacronym{peb}{PEB}{positioning error bound}
\newacronym{per}{PER}{packet error rate}
\newacronym{pet}{PET}{privacy enhancing technology}
\newacronym{pg}{PG}{path gain}
\newacronym{pgd}{PGD}{proximal gradient descent}
\newacronym{phy}{PHY}{physical}
\newacronym{pki}{PKI}{public key infrastucture}
\newacronym{pl}{PL}{path loss}
\newacronym{pla}{PLA}{physically large array}
\newacronym{pll}{PLL}{phase-locked loop}
\newacronym[plural=PMs,firstplural=person months (PMs)]{pm}{PM}{person month}
\newacronym{pmf}{PMF}{polymer microwave fiber}
\newacronym{pmu}{PMU}{power management unit}
\newacronym{pn}{PN}{pseudo-noise}
\newacronym{po}{PO}{phase offset}
\newacronym{poc}{PoC}{proof of concept}
\newacronym{poe}{PoE}{power-over-Ethernet}
\newacronym{pps}{1PPS}{pulse per second}
\newacronym{pr}{PR}{phase reversal}
\newacronym{prb}{PRB}{Physical Resource Block}
\newacronym{prbs}{PRBs}{Physical Resource Blocks}
\newacronym{prs}{PRS}{Peripheral Reflex System}
\newacronym{ps}{PS}{Processing System}
\newacronym{psd}{PSD}{power spectral density}
\newacronym{pse}{PSE}{power sourcing equipment}
\newacronym{psk}{PSK}{phase shift keying}
\newacronym{psm}{PSM}{power saving mode}
\newacronym{pss}{PSS}{primary synchronisation signal}
\newacronym{ptp}{PTP}{precision-time protocol}
\newacronym{ptrs}{PTRS}{Phase-Tracking Reference Signals}
\newacronym{ptw}{PTW}{paging time window}
\newacronym{pv}{PV}{photovoltaic}
\newacronym{pw}{PW}{planar wavefront}
\newacronym{pwm}{PWM}{pulse width modulation}
\newacronym{qam}{QAM}{quadrature amplitude modulation}
\newacronym{qos}{QoS}{quality-of-service}
\newacronym{qpsk}{QPSK}{quadrature phase-shift keying}
\newacronym{qrrls}{QR-RLS}{QR decomposition based recursive least squares}
\newacronym{quadriga}{QuaDRiGa}{QUAsi Deterministic RadIo channel GenerAtor}
\newacronym{ra}{RA}{Random Access}
\newacronym{ram}{RAM}{random-access memory}
\newacronym{ran}{RAN}{radio access network}
\newacronym{rar}{RAR}{Random Access Response}
\newacronym{rat}{RAT}{radio access technology}
\newacronym{rb}{Rb}{Rubidium}
\newacronym{rbs}{RBS}{radio base station}
\newacronym{rbw}{RBW}{resolution bandwidth}
\newacronym{rc}{RC}{raised-cosine}
\newacronym{rcs}{RCS}{radar cross section}
\newacronym{rdl}{RDL}{redistribution layer}
\newacronym{re}{RE}{radio element}
\newacronym{relu}{ReLU}{rectified linear unit}
\newacronym{rf}{RF}{radio frequency}
\newacronym{rfeh}{RFEH}{radio frequency energy harvesting}
\newacronym{rfic}{RFIC}{radio-frequency integrated circuit}
\newacronym{rfid}{RFID}{radio frequency identification}
\newacronym{rfpt}{RFPT}{radio frequency power transfer}
\newacronym{rfsoc}{RFSoC}{Radio Frequency System-on-Chip}
\newacronym{rir}{RIR}{room impulse response}
\newacronym{ris}{RIS}{reflective intelligent surface}
\newacronym{rllmtc}{RLLMTC}{reliable low latency machine type communication}
\newacronym{rls}{RLS}{recursive least squares}
\newacronym{rms}{RMS}{root-mean-square}
\newacronym{rmse}{RMSE}{root-mean-square error}
\newacronym{rmt}{RMT}{random matrix theory}
\newacronym{rof}{RoF}{radio-over-fiber}
\newacronym{ros}{ROS}{robot operating system}
\newacronym{rpi}{RPi}{Raspberry Pi}
\newacronym{rrc}{RRC}{Radio Resource Connection}
\newacronym{rreq}{RREQ}{route request packet}
\newacronym{rsrp}{RSRP}{Reference Signals Received Power}
\newacronym{rsrq}{RSRQ}{Reference Signal Received Quality}
\newacronym{rss}{RSS}{received signal strength}
\newacronym{rssi}{RSSI}{received signal strength indicator}
\newacronym{rtc}{RTC}{real time clock}
\newacronym{rtf}{RTF}{reader talks first}
\newacronym{rtk}{RTK}{real time kinematics}
\newacronym{rts}{RTS}{ray tracing simulator}
\newacronym{ru}{RU}{Radio Unit}
\newacronym{rv}{RV}{random variable}
\newacronym{rw}{RW}{RadioWeaves}
\newacronym{rx}{RX}{receiver}
\newacronym{rzf}{RZF}{regularized zero forcing}
\newacronym{s-parameter}{S-parameter}{scattering parameter}
\newacronym{sa}{SA}{synchronization anchor}
\newacronym{sa5}{SA}{Stand Alone}
\newacronym{sar}{SAR}{specific absorption rate}
\newacronym{sbl}{SBL}{sparse Bayesian learning}
\newacronym{sc}{SC}{single carrier}
\newacronym{scfde}{SC-FDE}{single-carrier modulation with frequency-domain-equalization}
\newacronym{scfdma}{SCFDMA}{single-carrier frequency division multiple access}
\newacronym{scs}{SCS}{sub-carrier spacing}
\newacronym{sdg}{SDG}{Sustainable Development Goal}
\newacronym{sdm}{SDM}{sigma-delta modulator}
\newacronym{sdma}{SDMA}{spatial-division multiple access}
\newacronym{sdn}{SDN}{software-defined network}
\newacronym{sdof}{SDoF}{sigma-delta over fiber}
\newacronym{sdr}{SDR}{software-defined radio}
\newacronym{se}{SE}{spectral efficiency}
\newacronym{sei}{SEI}{specific emitter identification}
\newacronym{ser}{SER}{symbol-error rate}
\newacronym{sf}{SF}{spreading factor}
\newacronym{sfn}{SFN}{single frequency network}
\newacronym{sfp}{SFP}{small form-factor pluggable}
\newacronym{sha}{SHA}{Secure Hash Algorithm}
\newacronym{SigMF}{SigMF}{Signal Metadata Format}
\newacronym{simo}{SIMO}{single-input multiple-output}
\newacronym{sinr}{SINR}{signal-to-interference-plus-noise ratio}
\newacronym{siso}{SISO}{single-input single-output}
\newacronym{slam}{SLAM}{simultaneous localization and mapping}
\newacronym{slc}{SLC}{spatial leakage suppression}
\newacronym{slerp}{SLERP}{spherical linear interpolation}
\newacronym{sma}{SMA}{SubMiniature version A}
\newacronym{smc}{SMC}{specular multipath component}
\newacronym{smps}{SMPS}{switched mode power supply}
\newacronym{sndr}{SNDR}{signal-to-noise-and-distortion ratio}
\newacronym{snidr}{SNIDR}{signal-to-noise-and-interference-and-distortion ratio}
\newacronym{snir}{SNIR}{signal-to-interference-plus-noise ratio}
\newacronym{snr}{SNR}{signal-to-noise ratio}
\newacronym{soc}{SoC}{state of charge}
\newacronym{SoC}{SOC}{System on Chip}
\newacronym{sp}{SP}{service point}
\newacronym{spdt}{SPDT}{single pole double throw}
\newacronym{spi}{SPI}{Serial Peripheral Interface}
\newacronym{spst}{SPST}{single pole single throw}
\newacronym{sram}{SRAM}{static random-access memory}
\newacronym{srd}{SRD}{short-range device}
\newacronym{srls}{SRLS}{standard recursive least squares}
\newacronym{srs}{SRS}{Sounding Reference Signal}
\newacronym{ssb}{SSB}{synchronisation signal block}
\newacronym{ssd}{SSD}{solid state drive}
\newacronym{ssq}{SSQ}{simulator sickness questionnaire}
\newacronym{steam}{STEAM}{science, technology, engineering, the arts, and mathematics}
\newacronym{svd}{SVD}{singular value decomposition}
\newacronym{sw}{SW}{spherical wavefront}
\newacronym{swipt}{SWIPT}{simultaneous wireless information and power transfer}
\newacronym{synce}{SyncE}{Synchronous Ethernet}
\newacronym{ta}{TA}{timing advance}
\newacronym{tau}{TAU}{tracking area update}
\newacronym{tcer}{TCER}{transported to consumed energy ratio}
\newacronym{tcp}{TCP}{Transmission Control Protocol}
\newacronym{tcxo}{TCXO}{temperature compensated crystal oscillator}
\newacronym{tdce}{TD-CE}{time-domain compression and expansion}
\newacronym{tdd}{TDD}{time division duplexing}
\newacronym{tdma}{TDMA}{time division-multiple access}
\newacronym{tdoa}{TDOA}{time-difference-of-arrival}
\newacronym{thz}{THz}{Terahertz}
\newacronym{tl}{TL}{transfer learning}
\newacronym{to}{TO}{timing offset}
\newacronym{toa}{TOA}{time-of-arrival}
\newacronym{tof}{ToF}{time-of-flight}
\newacronym{tosm}{TOSM}{through-open-short-match}
\newacronym{tpms}{TPMS}{Tire-Pressure Monitoring System}
\newacronym{trl}{TRL}{technology readyness level}
\newacronym{trp}{TRP}{Transmission Reception Point}
\newacronym{tsn}{TSN}{time-sensitive networking}
\newacronym{ttf}{TTF}{tag talks first}
\newacronym{ttff}{TTFF}{Time To First Fix}
\newacronym{tti}{TTI}{transmission time interval}
\newacronym{ttm}{TTM}{time to market}
\newacronym{ttn}{TTN}{The Things Network}
\newacronym{tx}{TX}{transmitter}
\newacronym{uart}{UART}{Universal Asynchronous Receiver/Transmitter}
\newacronym{uav}{UAV}{unmanned aerial vehicle}
\newacronym{uc}{UC}{use case}
\newacronym{ucie}{UCIe}{Universal Chiplet Interconnect Express}
\newacronym{udp}{UDP}{User Datagram Protocol}
\newacronym{ue}{UE}{user equipment}
\newacronym{ugv}{UGV}{unmanned ground vehicle}
\newacronym{uhd}{UHD}{USRP hardware driver}
\newacronym{uhf}{UHF}{ultra-high frequency}
\newacronym{ul}{UL}{uplink}
\newacronym{ula}{ULA}{uniform linear array}
\newacronym{uMIMO}{$\mu$-MIMO}{ultra-massive Multiple-Input Multiple-Output}
\newacronym{ummtc}{umMTC}{ultra massive machine type communication}
\newacronym{un}{UN}{United Nations}
\newacronym{unow}{uNOW}{unified non-orthogonal waveform}
\newacronym{upa}{UPA}{uniform planar array}
\newacronym{ura}{URA}{uniform rectangular array}
\newacronym{urllc}{URLLC}{ultra-reliable low-latency communications}
\newacronym{usrp}{USRP}{universal software radio peripheral}
\newacronym{uv}{UV}{unmanned vehicle}
\newacronym{uw}{UW}{unique word}
\newacronym{uwb}{UWB}{ultrawideband}
\newacronym{uwofdm}{UW-DFT-s-OFDM}{unique word DFT-s-OFDM}
\newacronym{v2v}{V2V}{vehicle-to-vehicle}
\newacronym{vco}{VCO}{voltage-controlled oscillator}
\newacronym{vep}{VEP}{virtual edge platform}
\newacronym{vlc}{VLC}{visible light communication}
\newacronym{vlp}{VLP}{visible light positioning}
\newacronym{vna}{VNA}{vector network analyzer}
\newacronym{voc}{VOC}{Voltatile Organic Compound}
\newacronym{votable}{VOTable}{Virtual Observatory Table}
\newacronym{vr}{VR}{virtual reality}
\newacronym{vuca}{VUCA}{volatile, uncertain, complex and ambiguous}
\newacronym{wb}{WB}{wideband}
\newacronym{wban}{WBAN}{wireless body area network}
\newacronym{wd}{WD}{Wireless distance}
\newacronym{wimax}{WiMAX}{Worldwide Interoperability for Microwave Access}
\newacronym{wlan}{WLAN}{wireless LAN}
\newacronym[plural=WPs,firstplural=work packages (WPs)]{wp}{WP}{work package}
\newacronym{wpt}{WPT}{wireless power transfer}
\newacronym{wr}{WR}{White Rabbit}
\newacronym{wrsn}{WRSN}{wireless rechargeable sensor network}
\newacronym{wsn}{WSN}{wireless sensor network}
\newacronym{xets}{XETS}{cross exponentially tapered slot}
\newacronym{xlmimo}{XL-MIMO}{extremely large-scale MIMO}
\newacronym{xr}{XR}{extended reality}
\newacronym{z3ro}{Z3RO}{zero third-order distortion}
\newacronym{zf}{ZF}{zero-forcing}
\newacronym{zmcscg}{ZMCSCG}{zero mean circularly symmetric complex Gaussian}
\newacronym{zmq}{ZMQ}{ZeroMQ}
\newacronym{ztofdm}{ZT-DFT-s-OFDM}{zero-tail DFT-s-OFDM}

\makeatletter
\def\myColorList{LimeGreen,BrickRed,Fuchsia,Bittersweet,YellowOrange, YellowGreen, WildStrawberry, Fuchsia, RedViolet, Periwinkle, PineGreen, OrangeRed, RawSienna, Turquoise, Aquamarine, BlueViolet, BurntOrange, DarkOrchid, TealBlue, Violet, RoyalPurple,%
LimeGreen,BrickRed,Fuchsia,Bittersweet,YellowOrange, YellowGreen, WildStrawberry, Fuchsia, RedViolet, Periwinkle, PineGreen, OrangeRed, RawSienna, Turquoise, Aquamarine, BlueViolet, BurntOrange, DarkOrchid, TealBlue, Violet, RoyalPurple}
\readlist\ColorList\myColorList

\definecolor{defaultColor}{RGB}{112,45,128}
\NewDocumentCommand{\roundLabel}{O{defaultColor} m}{%
\tikzexternaldisable
    \tikz[baseline=(char.base)]{
        \node[rectangle, rounded corners=0.65mm, inner sep=0.65mm, fill=#1, draw=#1, text=white, font=\itshape](char) {#2};
    }%
\tikzexternalenable
}

\usepackage[capitalize]{cleveref}

\begin{document}
\glsdisablehyper

\title{GNN-based Precoder Design and Fine-tuning for Cell-free Massive MIMO with Real-world CSI}

\author{
    \IEEEauthorblockN{Tianzheng Miao, Thomas Feys, Gilles Callebaut, Jarne Van Mulders, Emanuele Peschiera,\\ Md Arifur Rahman, François Rottenberg}
    \IEEEauthorblockA{
        \textit{KU Leuven, ESAT-WaveCore, Ghent Technology Campus, Ghent, Belgium}\\
        \textit{Research and Innovation Department, IS-Wireless, Piaseczno, Poland}
    }
    \thanks{This work was supported by the European Union's Horizon 2022 research and innovation program under Grant Agreement No 101120332. \textit{(Corresponding author: Tianzheng Miao)}}
}

\maketitle

\vspace{-1.5cm}

\begin{abstract}
\Gls{cfmmimo} has emerged as a promising paradigm for delivering uniformly high-quality coverage in future wireless networks. To address the inherent challenges of precoding in such distributed systems, recent studies have explored the use of \gls{gnn}-based methods, using their powerful representation capabilities. However, these approaches have predominantly been trained and validated on synthetic datasets, leaving their generalizability to real-world propagation environments largely unverified. In this work, we initially pre-train the \gls{gnn} using simulated \gls{csi} data, which incorporates standard propagation models and small-scale Rayleigh fading. Subsequently, we fine-tune the model on real-world \gls{csi} measurements collected from a physical testbed equipped with distributed \glspl{ap}. To balance the retention of pre-trained features with adaptation to real-world conditions, we adopt a layer-freezing strategy during fine-tuning, wherein several \gls{gnn} layers are frozen and only the later layers remain trainable. Numerical results demonstrate that the fine-tuned \gls{gnn} significantly outperforms the pre-trained model, achieving an approximate $\mathbf{8.2}$ bits per channel use gain at \SI{20}{dB} \gls{snr}, corresponding to a \SI{15.7}{\percent} improvement. These findings highlight the critical role of transfer learning and underscore the potential of \gls{gnn}-based precoding techniques to effectively generalize from synthetic to real-world wireless environments.
\end{abstract}

\glsresetall
\section{Introduction}

\Gls{cfmmimo} has emerged as a promising technology for future wireless communication systems, characterized by numerous distributed \glspl{ap} interconnected with one or more \glspl{cpu}. This architecture enables cooperative service to all users within a given area via joint signal encoding and decoding based on users’ \gls{csi}, which is either locally estimated at each \gls{ap} or centrally aggregated~\cite{ref1}. However, as the network scales up, efficiently managing the increasing volume of \gls{csi} and enabling robust precoding across diverse deployment scenarios remain key challenges. This highlights the need for scalable and generalizable precoding methods that can adapt to practical propagation conditions beyond idealized assumptions.

To overcome this challenge, recent advancements in learning-based methods, particularly those using \glspl{gnn}, have shown significant potential by exploiting the underlying topological structure of wireless networks~\cite{ref5}. For example, \gls{gnn}-based methods have successfully facilitated implicit channel estimation in \gls{ris} systems by directly mapping pilot signals to beamforming configurations through permutation-invariant architectures~\cite{ref8}. Additionally, \gls{gnn} approaches have demonstrated adaptability against various sources of signal degradation. Recent studies indicate superior performance of \gls{gnn}-based precoding compared to traditional methods, for instance in scenarios involving nonlinear power amplifier distortions~\cite{ref4}. 

However, despite the promising performance of existing \gls{gnn}-based precoding approaches, most of them have been trained and evaluated on synthetic datasets. This is primarily due to the high cost and complexity involved in collecting large-scale, high-quality real-world channel measurements~\cite{ref10}. While synthetic data enables rapid experimentation, it often fails to capture the rich variability and hardware impairments present in practical deployments, limiting the generalization capability of learned models~\cite{ref6}. Ideally, training directly on real measurement data would improve robustness and deployment readiness. However, the limited availability of such data motivates the exploration of more efficient learning strategies. In this context, \gls{tl}, which leverages knowledge gained from prior tasks or domains to improve learning efficiency in new ones~\cite{ref7}, has emerged as a promising approach to bridge the sim-to-real gap. A typical example is to pre-train a model on synthetic data and then fine-tune it using a small amount of real-world measurements~\cite{ref11}. This motivates our work to investigate how \gls{gnn}-based precoding models can be effectively adapted to realistic propagation scenarios using limited real measurements.

This work proposes a novel \gls{gnn}-based precoding framework designed for \gls{cfmmimo} systems. In the first stage, a \gls{gnn} is designed to learn a mapping from channel matrices to precoding matrices in an unsupervised manner. The model is initially pretrained on a large-size synthetic dataset, where channel matrices are generated using standard geometric propagation models combined with small-scale Rayleigh fading.
In the second stage, the pretrained \gls{gnn} is fine-tuned using real-world \gls{csi} measurements collected from the Techtile testbed, enabling the model to adapt to practical channel conditions. To balance the retention of useful knowledge acquired during synthetic pretraining with the need to adapt to domain-specific distributions of real-world \gls{csi}, a strategic layer-freezing scheme is introduced. In this scheme, selected layers of the network are frozen during fine-tuning to preserve generalizable representations while allowing deeper layers to specialize to the target domain.
The performance of the fine-tuned model is evaluated on both synthetic and real \gls{csi} datasets, and benchmarked against conventional precoding techniques. Numerical results show that fine-tuning leads to a substantial performance gain—improving the sum-rate by approximately \SI{8.2}{bits/channel,use}, or about \SI{15.7}{\percent}—thus demonstrating the effectiveness of transfer learning in enabling practical generalization for real-world deployment.

\textit{Notations:} Boldface lowercase and uppercase letters denote vectors and matrices, respectively. The operators $(\cdot)^\mathrm{T}$ and $(\cdot)^H$ indicate matrix transpose and conjugate transpose operations, respectively. The trace of a matrix is given by $\operatorname{Tr}(\cdot)$. The set of complex numbers is represented by $\mathbb{C}$.
\section{System Model}
\begin{figure}[htpb]
\centering
\includegraphics[width=0.4\textwidth]{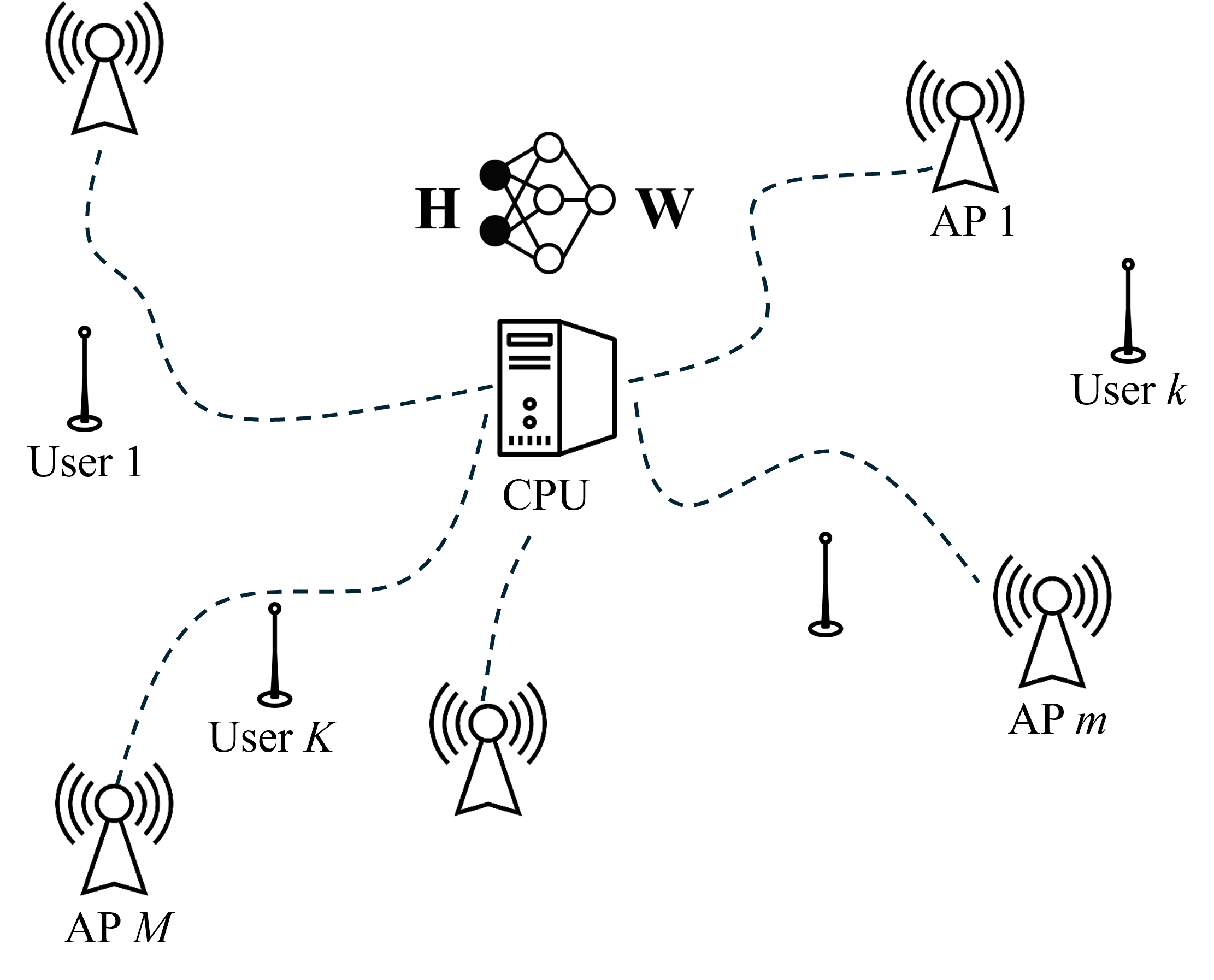}
\caption{System model of the considered \gls{cfmmimo} network. Distributed \glspl{ap} serve multiple single-antenna \glspl{ue} with coordination by a \gls{cpu}, which handles joint signal processing.}
\label{fig:6}
\end{figure}

As illustrated in~\cref{fig:6}, we consider a \gls{cfmmimo} network comprising \( M \) single-antenna \glspl{ap} and \( K \) single-antenna \glspl{ue}. Let \( m = 1, 2, \dots, M \) and \( k = 1, 2, \dots, K \) index the \glspl{ap} and \glspl{ue}, respectively. All \glspl{ap} are equipped with a single isotropic antenna and connected via fronthaul links to a \gls{cpu}, which performs centralized channel estimation and precoding based on global \gls{csi}.

In this work, we consider a fully centralized downlink scenario in which the \glspl{ap} are randomly located within a specific geographical area, and each \gls{ue} is simultaneously served by all \glspl{ap} with $M > K$. 
The channel between the \glspl{ap} and the \glspl{ue} is represented by the channel matrix $\mathbf{G} \in \mathbb{C}^{K \times M}$, where the channel coefficient between \gls{ap} $m$ and \gls{ue} $k$ is expressed as
\begin{equation}
g_{m,k} = \sqrt{\beta_{m, k}} h_{m, k}
\end{equation}
where $\beta_{m,k}$ is the large-scale fading coefficient, capturing path loss effects, and $h_{m,k}$ is the small-scale fading coefficient. Specifically, the path loss (in dB) is modeled according to the Indoor Hotspot (InH) Non-Line-of-Sight (NLOS) scenario~\cite{ref3}
\begin{equation}
\beta_{m,k} = 32.4 + 31.9\log_{10}(d_{m, k}) + 20\log_{10}(f_c)
\label{equ1}
\end{equation}
where $d_{m k}$ denotes the distance in meters between \gls{ap} $m$ and \gls{ue} $k$, and $f_c$ is the carrier frequency in GHz. The small scale fading coefficients are \gls{iid}, where $h_{m,k} \sim \mathcal{CN}(0,1)$ meaning that each coefficient follows a complex Gaussian distribution. To simultaneously serve all \glspl{ue}, each \gls{ap} participates in linear precoding. Let $\mathbf{W}=\left[\mathbf{w}_1, \mathbf{w}_2, \ldots, \mathbf{w}_K\right] \in \mathbb{C}^{M \times K}$ denote the precoding matrix, where $\mathbf{w}_k \in \mathbb{C}^{M}$ represents the precoding vector towards \gls{ue} $k$ from all \glspl{ap}. Accordingly, the received signal at \gls{ue} $k$ is thus given by

\begin{equation}
y_k = \mathbf{g}_k^\mathrm{T} \mathbf{w}_k s_k +\sum_{l=1,l\neq k}^K \mathbf{g}_k^\mathrm{T}\mathbf{w}_l s_l+n_k
\end{equation}
where \( \mathbf{g}_k \in \mathbb{C}^{M} \) denotes the channel vector between \gls{ue} \(k\) and the \glspl{ap}, and \( n_k \sim \mathcal{CN}(0, \sigma^2) \) is additive white Gaussian noise at \gls{ue} \(k\). The transmitted symbols \( s_k \sim \mathcal{CN}(0, 1) \) are independent and identically distributed (i.i.d.) complex Gaussian random variables, uncorrelated across different \glspl{ue}.
Accordingly, the \gls{sinr} at \gls{ue} $k$ is calculated as
$$
\operatorname{SINR}_k=\frac{\left|\mathbf{g}_k^{\mathrm{T}} \mathbf{w}_k\right|^2}{\sum_{l=1, l \neq k}^K\left|\mathbf{g}_k^{\mathrm{T}} \mathbf{w}_l\right|^2+\sigma^2}.
$$

Thus, the sum rate of the system, combining the individual user rates, can be expressed as
\begin{equation}
R_{\text{sum}} = \sum_{k=1}^{K} R_k = \sum_{k=1}^{K} \log_2\left(1 + \operatorname{SINR}_k\right).
\label{equ:1}
\end{equation}

\section{\gls{gnn}-based Precoder Design}

\begin{figure}[htbp]
    \includegraphics[width=0.8\linewidth]{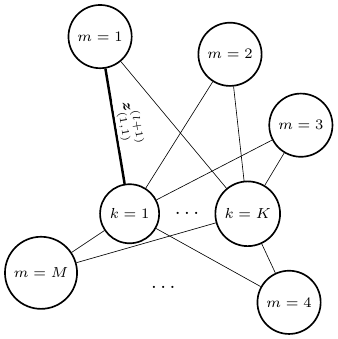}
    \centering
    \caption{Illustration of the graph representing the cell-free system}
    \label{fig:gnn}
\end{figure}

In the following, we describe the design and fine-tuning strategy of the proposed \gls{gnn}-based precoder, shown in Fig.~\ref{fig:gnn}. The objective of our neural network is to learn a mapping from the channel matrix $\mathbf{G}$ to the corresponding precoding matrix $\mathbf{W}$. Our \gls{gnn} architecture comprises \num{8} layers, each executing message-passing operations on the graph-structured representation of the wireless network. 

\begin{figure*}[htpb]
    \centering
    \includegraphics[width=0.8\textwidth]{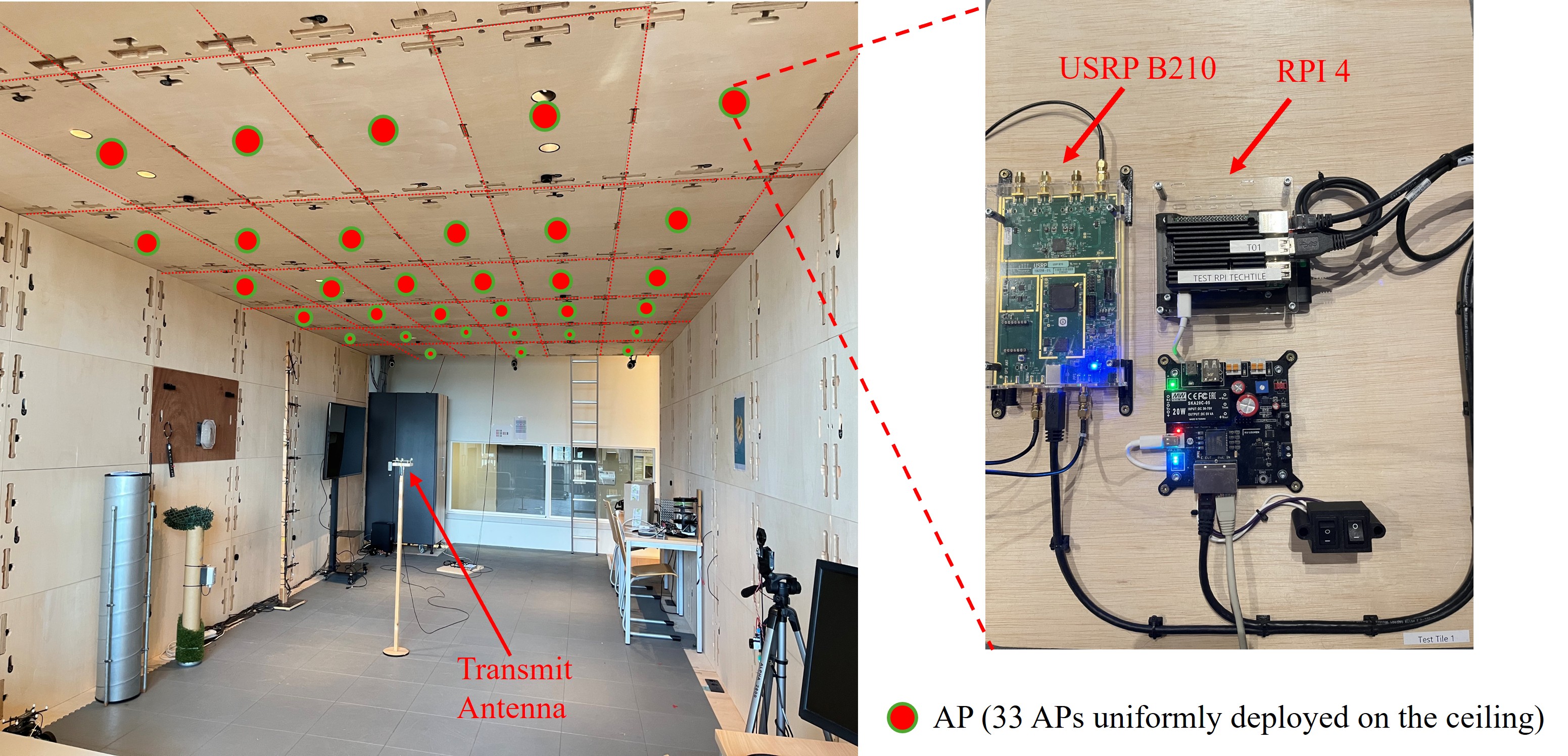}
    \caption{Testbed setup (\textit{Left:} Techtile environment with \num{33} \glspl{ap} on the ceiling and \gls{ue} at floor. \textit{Right:} Illustration of the hardware setup for each \gls{ap}, deployed on the backside of the ceiling planks.)}
    \label{fig:4}
\end{figure*}

Inspired by recent advances in the literature~\cite{ref4}, we adopt an edge-centric representation to better capture the wireless propagation characteristics, as edges naturally correspond to the communication channels between antennas at the \glspl{ap} and the \glspl{ue}. In this formulation, the \gls{cfmmimo} network is modelled as a bipartite graph $\mathcal{G}=(\mathcal{V}, \mathcal{E})$, where the vertex set $\mathcal{V}$ includes nodes representing the \glspl{ap} and the \glspl{ue}. The edge set $\mathcal{E}$ encodes the wireless links, with \gls{csi} serving as edge attributes. 

Each \gls{gnn} layer updates its edge representations through message passing, aggregating information from neighboring nodes and edges according to
\begin{equation}
\mathbf{z}_{(m,k)}^{(l+1)} = \text{UPDATE}\left(\mathbf{z}_{(m,k)}^{(l)}, \mathbf{m}_{(m)}^{(l)}, \mathbf{m}_{(k)}^{(l)}\right)
\end{equation}
where $\mathbf{z}_{(m,k)}^{(l)}$ denotes the representation of the edge connecting \gls{ap} $m$ and \gls{ue} $k$ at layer $l$, and $\mathbf{m}_{(m)}^{(l)}$, $\mathbf{m}_{(k)}^{(l)}$ represent aggregated messages from neighboring edges. Node messages are computed using aggregation functions as follows
\begin{equation}
\mathbf{m}_{(v)}^{(l)} = \text{AGGREGATE}\left({\mathbf{z}_{(v,u)}^{(l)} \mid u \in \mathcal{N}(v)}\right).
\end{equation}
In our model, the AGGREGATE function is defined as the element-wise mean over incoming edge features
\begin{equation}
\mathbf{m}{(v)}^{(l)} = \frac{1}{|\mathcal{N}(v)|} \sum_{u \in \mathcal{N}(v)} \mathbf{z}_{(v,u)}^{(l)}.
\end{equation}

The UPDATE function linearly combines the current edge embedding with the aggregated messages from both endpoint nodes
\begin{equation}
\mathbf{z}_{(m,k)}^{(l+1)} = \sigma\left( \mathbf{W}_{\text{edge}}^{(l)} \mathbf{z}_{(m,k)}^{(l)} + \mathbf{W}_m^{(l)} \mathbf{m}_{(m)}^{(l)} + \mathbf{W}_k^{(l)} \mathbf{m}_{(k)}^{(l)} \right),
\end{equation}
where $\sigma(\cdot)$ denotes a LeakyReLU activation function, and $\mathbf{W}_{\text{edge}}^{(l)}$, $\mathbf{W}_m^{(l)}$, and $\mathbf{W}_k^{(l)}$ are trainable weight matrices at layer $l$.

To adhere to transmit power constraints inherent in wireless communication systems, a power normalization step is integrated, defined by
\begin{equation}
\mathbf{W}^{\text{norm}} = \alpha \mathbf{W}, \quad \text{with} \quad \alpha = \sqrt{P_\mathrm{T} / \text{Tr}(\mathbf{W}\mathbf{W}^H)}
\end{equation}
where $P_\mathrm{T}$ represents the total available transmit power.
The proposed \gls{gnn}-based precoder is trained in an unsupervised manner with the objective of maximizing the sum rate in~\cref{equ:1}. 

\section{Transfer Learning with Real-World Data}

In this study, we employ \gls{tl} to transfer knowledge learned from synthetic data to real-world scenarios. Specifically, after an initial pretraining phase on large-scale synthetic datasets, we fine-tune the pretrained \gls{gnn} model using a limited-size real-world dataset. This fine-tuning process is also conducted in an unsupervised manner.

\subsection{Data Collection and Dataset Preparation}
To assess the model's ability to generalize from simulated to real-world environments, we collected real \gls{csi} data using the Techtile testbed~\cite{ref2}, which emulates a physical cell-free massive MIMO system. As illustrated in Fig.~\ref{fig:4}, our experimental setup includes \num{33} ceiling-mounted \glspl{ap}, each implemented using a \gls{usrp} B210 software-defined radio and managed via a dedicated \gls{rpi} 4. Within the same space, a single \gls{ue} was used to collect channel data.

During the data acquisition process, the \gls{ue} transmits uplink pilot signals, which are coherently received by all \glspl{ap} to form the composite \gls{csi}. To ensure diverse spatial sampling and comprehensive coverage, the \gls{ue} was moved to a different position after each measurement. This procedure results in a dataset denoted as
$$
\mathcal{H} = \{ \mathbf{h}_1, \mathbf{h}_2, \dots, \mathbf{h}_{500} \}
$$
where each measurement vector \( \mathbf{h}_i \in \mathbb{C}^{1 \times M} \) represents the uplink \gls{csi} from a single-antenna user located at a specific spatial position to all \( M = 33\) \glspl{ap}. This ensures spatial variability across \num{500} unique positions.

To simulate a two-user communication scenario, we construct a new dataset by pairing the single-user channel vectors. For each unordered pair of distinct indices \( (i, j) \) where \( 1 \leq i < j \leq 500 \), we define a two-user sample as the tuple \( (\mathbf{h}_i, \mathbf{h}_j) \). The total number of such combinations is given by the binomial coefficient $\binom{500}{2}$, resulting in \num{124750} unique two-user channel instances. Each sample represents a realistic communication scenario in which two spatially separated users are jointly served by the same set of access points.

It is also possible to extend this setup to a four-user scenario by generating all unique unordered 4-tuples from the \num{500} single-user samples, with the total number of such combinations given by the binomial coefficient $\binom{500}{4}$. As this results in over \num{2.5} billion combinations, which is computationally infeasible to process in full. To balance the need for maintaining high-quality channel conditions with the requirement of controlling the dataset size for computational tractability and fair comparison across scenarios, we select the top \num{44} single-user samples based on channel strength \( \|\mathbf{h}_i\| \). This selection ensures a sufficiently large number of unique four-user combinations, as $\binom{44}{4} = 135751 \geq 124750$. From these, we randomly sample \num{124750} combinations without replacement to match the size of the two-user dataset.

For model training and evaluation, both the two-user and four-user datasets are partitioned into training, validation, and testing subsets, containing \num{80}\%, \num{10}\%, and \num{10}\% of the total samples, respectively. This split ensures sufficient diversity for effective model learning while maintaining reliable evaluation performance across unseen data.
The complete preprocessed dataset is publicly available at \href{https://kuleuven-my.sharepoint.com/:f:/g/personal/tianzheng_miao_kuleuven_be/ErQh2y6X6yxPj-w9WLlbVDIB52p2T0EXlAqj26fXwJBj7g?e=fOJr3W}{Real-world \gls{csi} Dataset}.

\subsection{Fine-Tuning Strategy}
\label{sec:3}
A key challenge in deploying models trained on synthetic data is the distribution mismatch between simulated and real-world environments, commonly known as covariate shift~\cite{ref13}. In our context, real \gls{csi} exhibits non-idealities and measurement noise absent from simulated data, often leading to degraded performance when directly applying a pretrained model.

To effectively adapt the pretrained model to real-world data while retaining its learned representations, we adopt a layer-freezing approach during fine-tuning. 
The underlying \gls{gnn} architecture consists of \num{8} layers, which allows for selective freezing at different depths of the network. Specifically, the first \textit{l} layers of the model are kept fixed (i.e., their parameters are not updated), and the remaining \textit{8-l} layers are retrained using the real-world dataset.

The proposed method aims to keep a balance between preserving useful knowledge acquired from simulation and enabling flexible adaptation to the domain shift introduced by real \gls{csi}. Furthermore, freezing early layers reduces the number of trainable parameters, which is particularly beneficial when only limited real training samples are available, as it helps mitigate overfitting while still allowing sufficient capacity in later layers to avoid underfitting.

To identify the optimal freezing point, we conducted an exhaustive evaluation across all possible values of $l \in\{0,1, \ldots, 8\}$. Notably, Freeze\_\textit{0} corresponds to full fine-tuning, where all layers are updated, while Freeze\_\textit{8} denotes a fully frozen model, effectively identical to the pretrained network without any adaptation. This layer-wise investigation enables us to assess the trade-off between stability and adaptability under real deployment conditions.

\section{Results}
This section presents the results of our experiments, focusing on identifying the optimal layer-freezing strategy and evaluating the performance of the proposed \gls{gnn}-precoder and fine-tuning network across various datasets. The \gls{snr} values shown on the x-axis are defined as $\mathrm{SNR}_{\mathrm{Tx}} = 10 \log_{10}(P_t / \sigma^2)$, where $\sigma^2$ denotes the noise variance. This represents the transmit-side \gls{snr} prior to any channel fading or path loss effects.
\subsection{Freezing Strategy and Fine-Tuning Performance}
\begin{figure}[htpb]
    \centering
    \includegraphics[width=0.9\linewidth]{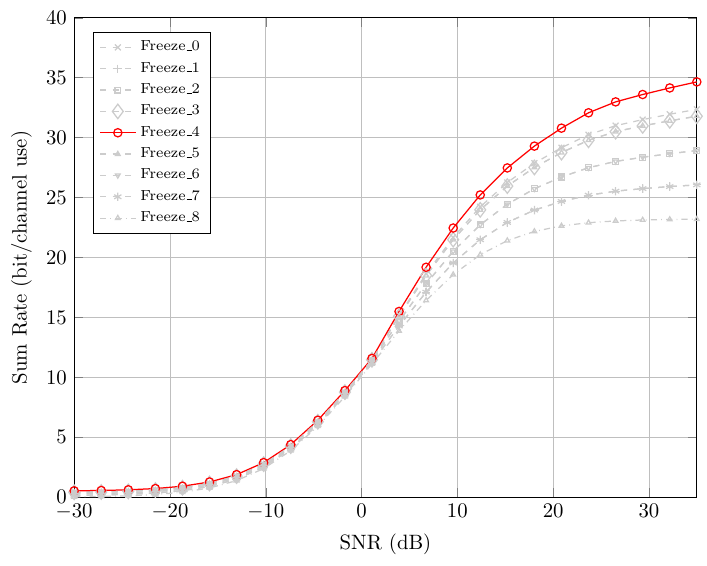}
    \caption{Comparison of different freezing strategies. Freeze\_\textit{l} indicates that the first \textit{l} layers of the network are frozen, while the remaining layers are retrained using the collected real-world dataset. Note that the evaluation was done on the real-world \gls{csi} dataset}
    \label{fig:3}
\end{figure}

The pretrained \gls{gnn} was initially trained using a synthetic dataset containing \num{500000} training samples and \num{50000} validation samples, followed by testing on \num{10000} samples. The training employed a learning rate of \num{0.005} over \num{20} epochs. Identical dataset sizes were used for both the two-user and four-user scenarios to ensure consistency in evaluation.

To determine the most effective freezing strategy for subsequent fine-tuning, we evaluated various configurations using real-world \gls{csi} data from a four-user scenario as the test set, with the learning rate kept consistent with that used during pretraining. As illustrated in Fig.~\ref{fig:3}, freezing the first \num{4} layers achieves the best performance. This configuration strikes a balance between retaining the pretrained model’s prior knowledge and maintaining sufficient adaptability to the new real-world dataset. Notably, the performance of Freeze\_{\textit{0}} ranks second, slightly outperforming Freeze\_{\textit{1}} and Freeze\_{\textit{7}}. In contrast, Freeze\_{\textit{8}} yields the poorest performance, as freezing all layers prevents the model from adapting to the new data. Based on these findings, we adopted the Freeze\_{\textit{4}} strategy for all subsequent fine-tuning experiments.

\subsection{Generalization on Real vs. Synthetic \Gls{csi}}
\begin{figure}[htpb]
\centering
\includegraphics[width=0.9\linewidth]{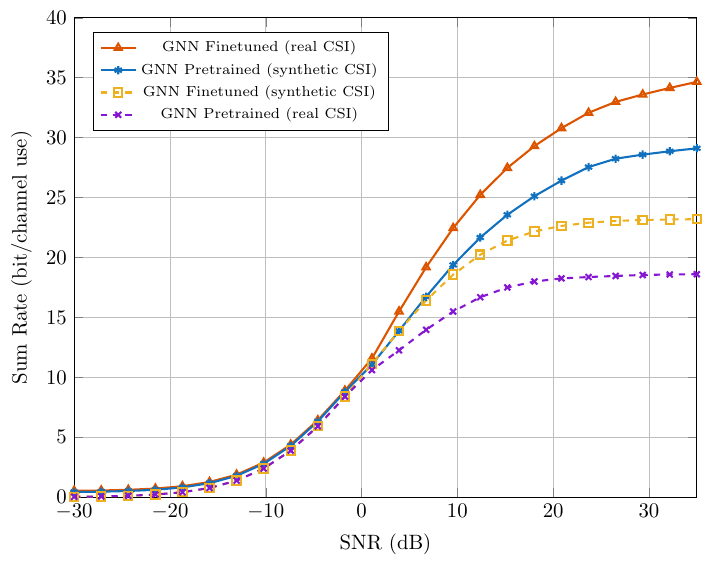}
\caption{Sum-rate performance comparison for different precoding methods with \num{4} \glspl{ue}, evaluated on real and synthetic \gls{csi}.}
\label{fig:2}\end{figure}

Figure~\ref{fig:2} illustrates the performance of the proposed \gls{gnn}-based precoding network under various conditions. Both real and synthetic \gls{csi} datasets were used to evaluate the pretrained and fine-tuned models across a range of transmit \gls{snr} values. The results show that the fine-tuned model consistently outperforms the pretrained one on the real-world dataset, while it underperforms on the synthetic dataset compared to the pretrained model. This outcome reflects the effect of \gls{tl} that fine-tuning enhances the model's ability to generalize to real data at the cost of some performance degradation on the synthetic domain. Moreover, the fine-tuned model performs better when evaluated on real \gls{csi} data than on synthetic data, demonstrating the benefits of domain adaptation through fine-tuning. Conversely, the pretrained model, which has not been exposed to real-world data, exhibits degraded performance when tested on the real \gls{csi} dataset, further highlighting the necessity of fine-tuning.

\subsection{Scalability of Fine-Tuned \gls{gnn} vs. Different Precoding Methods}
This subsection evaluates the scalability of the proposed fine-tuned GNN model by comparing its performance to traditional precoding schemes in both two-user and four-user settings. As baselines, we adopt \gls{zf} and \gls{cb}, two classical and widely used precoding methods~\cite{ref9}. These methods are defined as follows
\begin{equation}
    \mathbf{W}= \begin{cases}\alpha\mathbf{H}^H & \text {\gls{cb}} \\ \alpha\mathbf{H}^H\left(\mathbf{H}\mathbf{H}^H\right)^{-1} & \text {\gls{zf}}\end{cases}
\end{equation}
where \gls{zf} precoding is used as a performance reference in our experiment due to its theoretical ability to eliminate inter-user interference under ideal conditions.
\begin{figure}[htpb]
\centering
\begin{subfigure}{0.5\textwidth}
\centering
\includegraphics[width=0.9\linewidth]{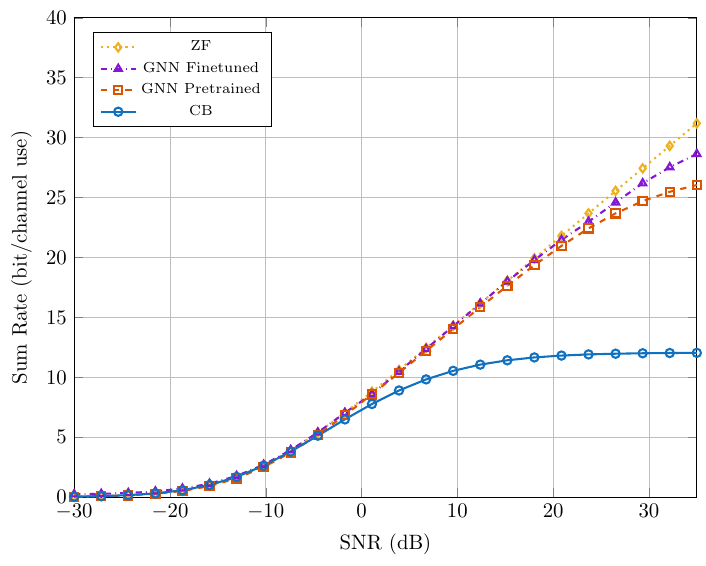}
\caption{Sum rate vs. \gls{snr} with \num{2} \glspl{ue}}
\label{fig:1a}
\end{subfigure}
\begin{subfigure}{0.5\textwidth}
\centering
\includegraphics[width=0.9\linewidth]{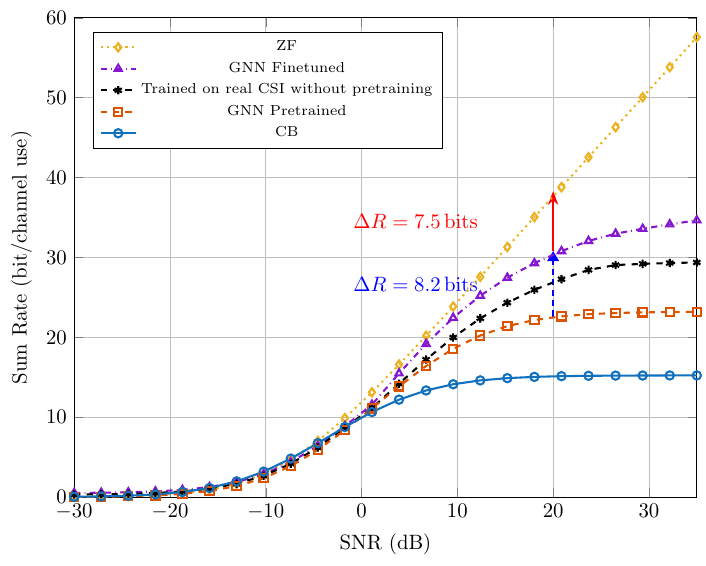}
\caption{Sum rate vs. \gls{snr} with \num{4} \glspl{ue}}
\label{fig:1b}
\end{subfigure}
\caption{Sum-rate performance of the \gls{cfmmimo} system with different numbers of \glspl{ue} ($M=33$) and precoding schemes. All methods are evaluated on real-world \gls{csi} data.}
\label{fig:1}
\end{figure}

As shown in Fig.~\ref{fig:1a}, the pretrained \gls{gnn} surpasses \gls{cb} and approaches the performance of \gls{zf} up to \SI{20}{dB} \gls{snr}. Beyond this point, its performance saturates. This limitation is attributed to the fact that the model was trained at a fixed \gls{snr} level, which reduces its ability to generalize to higher \gls{snr} regimes not seen during training. In contrast, \gls{zf} maintains robust performance by analytically computing the pseudo-inverse of the channel matrix.

In the four-user case (Fig.~\ref{fig:1b}), the pretrained \gls{gnn} still outperforms \gls{cb}, but lags significantly behind \gls{zf}, with a performance gap of \SI{15.7}{bits/channel\,use}. After applying fine-tuning, this gap is reduced to \SI{7.5}{bits/channel\,use}, corresponding to a relative improvement of approximately \SI{15.7}{\percent}. The increased number of users introduces more complex inter-user interference, which challenges the generalization ability of the pretrained model. However, the fine-tuning process enables the GNN to adapt to these more difficult scenarios by learning from real-world interference patterns, thereby significantly narrowing the performance gap.

The performance of the fine-tuned \gls{gnn} compared with the baseline model trained directly on real \gls{csi} without pretraining is shown in~\cref{fig:1b}. It can be observed that the baseline model fails to match the performance of the fine-tuned \gls{gnn}. This performance gap is primarily due to the baseline being initialized with random weights, which can lead to convergence toward suboptimal solutions. In contrast, fine-tuning benefits from a favorable initialization derived from pretraining, which guides the optimization toward better local minima. Additionally, the inferior performance of the baseline can be attributed to the limited size of the real dataset, which may be insufficient to capture the full distributional characteristics of the underlying wireless channel.

\glsresetall
\section{Conclusion}
In this paper, we proposed a \gls{gnn}-based precoding framework for \gls{cfmmimo} systems, with a focus on enhancing its practical deployment through the application of \gls{tl}. The model was initially pretrained on a large-size synthetic \gls{csi} dataset generated using standard geometric propagation models combined with small-scale Rayleigh fading. To enable real-world applicability, the pretrained \gls{gnn} was subsequently fine-tuned using measured \gls{csi} data collected from a physical testbed featuring distributed \glspl{ap}.
To balance knowledge retention from pretraining and adaptability to domain shifts in real-world environments, a strategic layer-freezing scheme was employed during fine-tuning. Experimental results demonstrate that the fine-tuned \gls{gnn} achieves a notable performance improvement over the pretrained model, increasing the sum-rate by approximately \SI{8.2}{bits/channel\,use}, or about \SI{15.7}{\percent}.
\bibliographystyle{IEEEtran}
\bibliography{review}

\begin{thebibliography}{10}
\providecommand{\url}[1]{#1}
\csname url@samestyle\endcsname
\providecommand{\newblock}{\relax}
\providecommand{\bibinfo}[2]{#2}
\providecommand{\BIBentrySTDinterwordspacing}{\spaceskip=0pt\relax}
\providecommand{\BIBentryALTinterwordstretchfactor}{4}
\providecommand{\BIBentryALTinterwordspacing}{\spaceskip=\fontdimen2\font plus
\BIBentryALTinterwordstretchfactor\fontdimen3\font minus \fontdimen4\font\relax}
\providecommand{\BIBforeignlanguage}[2]{{%
\expandafter\ifx\csname l@#1\endcsname\relax
\typeout{** WARNING: IEEEtran.bst: No hyphenation pattern has been}%
\typeout{** loaded for the language `#1'. Using the pattern for}%
\typeout{** the default language instead.}%
\else
\language=\csname l@#1\endcsname
\fi
#2}}
\providecommand{\BIBdecl}{\relax}
\BIBdecl

\bibitem{ref1}
H.~Q. Ngo, A.~Ashikhmin, H.~Yang, E.~G. Larsson, and T.~L. Marzetta, ``Cell-free massive {MIMO} versus small cells,'' \emph{IEEE Transactions on Wireless Communications}, vol.~16, no.~3, pp. 1834--1850, 2017.

\bibitem{ref5}
M.~Lee, G.~Yu, H.~Dai, and G.~Y. Li, ``Graph neural networks meet wireless communications: Motivation, applications, and future directions,'' \emph{IEEE Wireless Communications}, vol.~29, no.~5, pp. 12--19, 2022.

\bibitem{ref8}
T.~Jiang, H.~V. Cheng, and W.~Yu, ``Learning to {{Reflect}} and to {{Beamform}} for {{Intelligent Reflecting Surface With Implicit Channel Estimation}},'' \emph{IEEE Journal on Selected Areas in Communications}, vol.~39, no.~7, pp. 1931--1945, Jul. 2021.

\bibitem{ref4}
T.~Feys, L.~Van~der Perre, and F.~Rottenberg, ``Toward energy-efficient massive {MIMO}: Graph neural network precoding for mitigating non-linear {PA} distortion,'' \emph{IEEE Transactions on Cognitive Communications and Networking}, vol.~11, no.~1, pp. 184--201, 2025.

\bibitem{ref10}
A.~Alkhateeb, ``{{DeepMIMO}}: {{A Generic Deep Learning Dataset}} for {{Millimeter Wave}} and {{Massive MIMO Applications}},'' Feb. 2019.

\bibitem{ref6}
Y.~Huangfu, J.~Wang, S.~Dai, R.~Li, J.~Wang, C.~Huang, and Z.~Zhang, ``{WAIR-D}: Wireless {AI} research dataset,'' \emph{arXiv preprint arXiv:2212.02159}, 2022.

\bibitem{ref7}
M.~Wang, Y.~Lin, Q.~Tian, and G.~Si, ``Transfer {{Learning Promotes 6G Wireless Communications}}: {{Recent Advances}} and {{Future Challenges}},'' \emph{IEEE Transactions on Reliability}, vol.~70, no.~2, pp. 790--807, Jun. 2021.

\bibitem{ref11}
------, ``Transfer {{Learning Promotes 6G Wireless Communications}}: {{Recent Advances}} and {{Future Challenges}},'' \emph{IEEE Transactions on Reliability}, vol.~70, no.~2, pp. 790--807, Jun. 2021.

\bibitem{ref3}
{3GPP}, ``Study on channel model for frequencies from 0.5 to 100 ghz (release 17),'' {3GPP}, Tech. Rep. TR 38.901 V17.0.0, 2022, (March 2022).

\bibitem{ref2}
G.~Callebaut, J.~Van~Mulders, G.~Ottoy, D.~Delabie, B.~Cox, N.~Stevens, and L.~Van~der Perre, ``Techtile: Open {6G R\&D} testbed for communication, positioning, sensing, {WPT} and federated learning,'' in \emph{Proc. Joint European Conference on Networks and Communications \& 6G Summit (EuCNC/6G Summit)}, 2022, pp. 417--422.

\bibitem{ref13}
J.~Quinonero-Candela, M.~Sugiyama, A.~Schwaighofer, and N.~D. Lawrence, \emph{Dataset Shift in Machine Learning}.\hskip 1em plus 0.5em minus 0.4em\relax MIT Press, 2008.

\bibitem{ref9}
E.~Bj{\"o}rnson, L.~Sanguinetti, J.~Hoydis, and M.~Debbah, ``Optimal {{Design}} of {{Energy-Efficient Multi-User MIMO Systems}}: {{Is Massive MIMO}} the {{Answer}}?'' \emph{IEEE Transactions on Wireless Communications}, vol.~14, no.~6, pp. 3059--3075, Jun. 2015.

\end{thebibliography}

\end{document}